\documentclass[aps,preprint,floatfix,showpacs,superscriptaddress]{revtex4}

\usepackage{graphicx}
\usepackage{amssymb}
\usepackage{bm}

\newcommand{\be}{\begin{equation}}
\newcommand{\ee}{\end{equation}}
\newcommand{\bel}[1]{\be\label{#1}}
\newcommand{\re}[1]{Eq.~(\ref{#1})}
\newcommand{\ds}{\displaystyle}
\newcommand{\ov}[1]{\overline{#1}}
\newcommand{\hsp}{\hspace*{1pt}}

\begin{document}

\title{
Longitudinal fluid dynamics for\\
ultrarelativistic heavy--ion collisions}

\author{L.M. Satarov}

\affiliation{Frankfurt Institute for Advanced Studies,
J.W.~Goethe University, Max--von--Laue--Str.~1,~D--60438 Frankfurt
am Main, Germany}

\affiliation{The Kurchatov Institute, Russian Research Center,
123182 Moscow, Russia}

\author{I.N.~Mishustin}

\affiliation{Frankfurt Institute for Advanced Studies,
J.W.~Goethe University, Max--von--Laue--Str.~1,~D--60438 Frankfurt
am Main, Germany}

\affiliation{The Kurchatov Institute, Russian Research Center,
123182 Moscow, Russia}

\author{A.V.~Merdeev}

\affiliation{The Kurchatov Institute, Russian Research Center,
123182 Moscow, Russia}

\author{H.~St\"ocker}

\affiliation{Frankfurt Institute for Advanced Studies,
J.W.~Goethe University, Max--von--Laue--Str.~1,~D--60438 Frankfurt
am Main, Germany}

\begin{abstract}
We develop a 1+1 dimensional hydrodynamical model for central
heavy--ion collisions at ultrarelativistic energies.
Deviations from Bjorken's scaling are taken into account by
implementing finite--size profiles for the initial energy density. The
calculated rapidity distributions of pions, kaons and antiprotons in
central Au+Au collisions at $\sqrt{s_{NN}}=200$\,GeV are compared with
experimental data of the BRAHMS Collaboration. The sensitivity of the
results to the choice of the equation of state, the parameters of initial
state and the freeze--out conditions is investigated.
Experimental constraints on the total energy of produced particles
are used to reduce the number of model parameters.
The best fits of experimental data are obtained for soft equations of state
and Gaussian--like initial profiles of the energy density.
It is found that initial energy densities required for
fitting experimental data decrease with increasing
critical temperature of the phase transition.
\end{abstract}


\maketitle

\section{Introduction}

High--energy heavy--ion collisions provide a unique tool for studying
properties of hot and dense strongly interacting matter in the
laboratory. The theoretical description of such collisions is often
done within the framework of a hydrodynamic approach.  This approach
opens the possibility to study the sensitivity of collision dynamics
and secondary particle distributions to the equation of state (EOS) of
the produced matter. The two most famous realizations of this approach,
which differ by the initial conditions, have been proposed by
Landau~\cite{Lan53} (full stopping) and Bjorken~\cite{Bjo83} (partial
transparency). In recent decades many versions of the hydrodynamic
model were developed ranging from
simplified 1+1~\cite{Mel58,Tar77,Mis83,Bla87,Esk98,Moh03} and 2+1 dimensional
models~\cite{Bla87,Kol99,Bas00,Per00,Kol01,Tea01} of the Landau or
Bjorken type to more sophisticated  3+1 dimensional models
\cite{Sto80,Ris95a,Non00,Hir02,Ham05,Non06}. One should also mention the
multi--fluid models
\cite{Ams78,Cla86,Bar87,Mis88,Kat93,Bra00,Ton03} which consider
the whole collision process including the nuclear interpenetration
stage. Recent theoretical investigations show that fluid--dynamical
models give a very good description of many observables at the SPS and
RHIC bombarding energies (see e.g. Ref.~\cite{Sto05}).

The 2+1 dimensional hydrodynamical models have been successfully
applied \mbox{\cite{Kol99,Bas00,Per00,Kol01,Tea01}} to describe the
$p_T$ distributions of mesons and their elliptic flow at midrapidity.
These models assume a boost--invariant expansion~\cite{Bjo83} of matter
in the longitudinal (beam) direction and, therefore, cannot explain
experimental data in a broad rapidity region, where strong deviations
from the scaling regime have been observed. More realistic 3+1
dimensional fluid--dynamical simulations have been already performed
for heavy--ion collisions at SPS and RHIC energies. But as a rule, the
authors of these models do not study the sensitivity of the results to
the choice of initial and final (freeze--out) stages. On the other
hand, it is not clear at present, which initial conditions,
Landau--like~\cite{Lan53} or Bjorken--like~\cite{Bjo83}, are more
appropriate for ultrarelativistic collisions.

Our main goal in this paper is
to see how well the fluid--dynamical approach can describe
the RHIC data on $\pi, K, \overline{p}$ distributions over a broad
rapidity interval, reported recently by the BRAHMS
Collaboration~\cite{Bea04,Bea05}. Within our approach we explicitly
impose a constraint on the total energy of the produced particles
which follows from these data. For our study we apply a simplified
version of the hydrodynamical model, dealing only with the longitudinal
dynamics of the fluid. This approach has as its limiting cases
the Landau and Bjorken models. We investigate the sensitivity
of the hadron rapidity spectra to the fluid's equation of state,
to the choice of initial state and freeze--out conditions.
Modification of these spectra due to the feeding
from resonance decays is also analyzed.
Special attention is paid to possible manifestations of the
deconfinement phase transition. In particular, we compare
the dynamical evolution of the fluid with and without the phase
transition.

\section{Formulation of the model}

\subsection{Dynamical equations}

Below we study the evolution of highly excited, and possibly
deconfined, strongly interacting matter produced in ultrarelativistic
heavy--ion collisions. It is assumed that after a certain thermalization
stage this evolution can be described by the ideal relativistic
hydrodynamics. The energy--momentum tensor is written in a standard
form\footnote{
Units with $\hbar=c=1$ are used throughout the paper.
}
\bel{enmt}
T^{\mu\nu}=(\epsilon+P)\hsp
 U^\mu U^\nu -P\hsp g^{\mu\nu},
\ee
where $\epsilon, P$ and $U^\mu$ are the rest--frame energy density,
pressure and the collective 4--velocity of the fluid.

We consider central collisions of equal nuclei disregarding the
effects of transverse collective expansion. It is convenient to
parametrize $U^\mu$ in terms of the longitudinal flow rapidity~$Y$ as
$U^\mu=(\cosh{Y},\bm{0},\sinh{Y})^\mu$. All calculations are performed
using the light--cone variables~\cite{Bjo83}, namely, the proper
time~$\tau$ and the space--time rapidity $\eta$\,, defined as
\bel{lcva}
\tau=\sqrt{t^2-z^2},\hspace*{5mm}\eta=\tanh^{-1}\left(\frac{z}{t}\right)=
\frac{1}{2}\ln{\frac{t+z}{t-z}}\,.
\ee
In these coordinates, the equations of relativistic hydrodynamics,
\mbox{$\partial_\nu T^{\mu\nu}=0$}, for an ideal
baryon--free fluid take the following form~\cite{Bel96}
\begin{eqnarray}
&&\left(\tau\frac{\partial}{\partial\tau}+
\tanh\hsp (Y-\eta)\hsp\frac{\partial}{\partial\eta}\right)\hsp\epsilon
+(\epsilon +P)\left(\tanh\hsp (Y-\eta)\hsp\tau\frac{\partial}{\partial\tau}+
\frac{\partial}{\partial\eta}\right) Y=0\hsp ,
\label{erf1}\\
&&(\epsilon +P)\left(\tau\frac{\partial}{\partial\tau}+\tanh\hsp
(Y-\eta)\hsp\frac{\partial}{\partial\eta}\right) Y
+\left(\tanh\hsp (Y-\eta)\hsp\tau\frac{\partial}{\partial\tau}+
\frac{\partial}{\partial\eta}\right) P=0\hsp .
\label{erf2}
\end{eqnarray}
To solve Eqs.~(\ref{erf1})--(\ref{erf2}), one needs to specify the EOS,
\mbox{$P=P\hsp (\epsilon)$}, and the initial profiles
$\epsilon\hsp (\tau_0,\eta)$ and $Y(\tau_0,\eta)$ at a time
$\tau=\tau_0$\, when the fluid may be considered as thermodynamically
equilibrated.

Following Ref.~\cite{Hir02}, we choose the initial conditions
for a finite-size fluid, generalizing the Bjorken scaling
conditions:
\bel{incd}
Y (\tau_0,\eta)=\eta,\hspace*{5mm}\epsilon\hsp (\tau_0,\eta)=\epsilon_0
\exp{\left[-\frac{(|\eta|-\eta_0)^2}{2\hsp\sigma^2}
\Theta(|\eta|-\eta_0)\right]}\,,
\ee
where $\Theta (x)\equiv (1+\textrm{sgn}\hsp x)/2$\,. The particular
choice $\eta_0=0$ corresponds to the pure Gaussian profile of the
energy density. At small $\sigma$ such a profile
can be similar to the Landau initial condition\footnote{
Within the Landau model~\cite{Lan53} $\sigma\propto \gamma^{-1}$
and $\epsilon_0\propto \gamma^2$ where $\gamma$ is the c.m.
Lorentz--factor of colliding nuclei.
}.
On the other hand, when $\sigma$ or $\eta_0$
tends to infinity, one gets the limiting case of the Bjorken scaling
solution.  Below we adopt the value $\tau_0=1$\,fm/{\it c}\,.

The numerical solution of Eqs.~(\ref{erf1})--(\ref{erf2}) is obtained
by using the relativistic version~\cite{Ris95b} of the flux--corrected
transport algorithm~\cite{Bor73}.

\subsection{Equation of state}

As well known, a deconfinement phase transition
is predicted by quantum chromodynamics (QCD).
This phase transition is implemented through
a bag--like EOS in the parametrization suggested
in Ref.~\cite{Tea01}. This EOS consists of three parts, denoted below by
indices $H, M$ and $Q$ corresponding, respectively, to the hadronic,
''mixed'' and quark--gluon phases. In the case of equilibrated
baryon--free matter the pressure $P$, energy density $\epsilon$ and entropy
density $s$ may be regarded as functions of the temperature only. The
hadronic phase consists of pions, kaons, meson resonances and
baryon--antibaryon pairs. It corresponds to the domain of low energy
densities, $\epsilon<\epsilon_H$, and temperatures, $T<T_H$\hsp .
The sound velocity, $c_s=\sqrt{dP/d\epsilon}$, is assumed to be
constant ($c_s=c_H$) in this phase:
\bel{heos}
P=c_H^2\epsilon\,,~T=T_H\left(\frac{\epsilon}
{\epsilon_H}\right)^{\frac{\ds c_H^2}{\ds 1+c_H^2}}~~~(\epsilon<\epsilon_H)\,.
\ee
The mixed phase corresponds to intermediate energy densities,
from $\epsilon_H$ up to~$\epsilon_Q$.
The following parametrization is used for this region:
\bel{meos}
P=c_M^2\epsilon-(1+c_M^2)\hsp B_M\,,~T=T_H\left(\frac{\epsilon-B_M}
{\epsilon_H-B_M}\right)^{\frac{\ds c_M^2}{\ds 1+c_M^2}}
~~~(\epsilon_H<\epsilon<\epsilon_Q)\hsp .
\ee
Here $B_M$ is the bag constant, determined from the condition of
continuity of~$P(\epsilon)$ at $\epsilon=\epsilon_H$\,. Due to the
small sound velocity $c_M$ (see Table~\ref{tab1}), both pressure and
temperature increase only weakly with $\epsilon$ in the mixed phase
region. The third, quark--gluon plasma region of the EOS corresponds to
energy densities above $\epsilon_Q$\,:
\bel{qeos}
P=c_Q^2\epsilon-(1+c_Q^2)\hsp B_Q\,,~T=T_Q\left(\frac{\epsilon-B_Q}
{\epsilon_Q-B_Q}\right)^{\frac{\ds c_Q^2}{\ds 1+c_Q^2}}
~~~(\epsilon>\epsilon_Q)\hsp .
\ee
Here $B_Q$ is the bag constant in the deconfined phase.
The corresponding formulae for the
entropy density are obtained from the thermodynamic relation
$s=(\epsilon+P)/T$\,. We use the sound velocities
$c_H^2, c_M^2, c_Q^2$ close to those used in Refs.~\cite{Kol99,Tea01}.
\begin{table}[htb!]
\caption{Parameters of EOS\hsp s with the deconfinement phase transition.}
\label{tab1}
\vspace*{2mm}
\begin{ruledtabular}
\begin{tabular}{cccccccccc}
& $\epsilon_H$ & $\epsilon_Q$ &
$c_H^2$ & $c_M^2$ & $c_Q^2$ & $T_H$ & $T_Q$ &
 $B_M$ & $B_Q$ \\[-2mm]
& (GeV/fm$^3$) & (GeV/fm$^3$) &&&& (MeV) & (MeV) & (MeV/fm$^3$) & (MeV/fm$^3$) \\
\hline
EOS--I & 0.45 & 1.65 &0.15~&0.02~&1/3~& 165 & 169 & $-57.4$ & 344\\
EOS--II & 0.79 & 2.90 &0.15~&0.02~&1/3~& 190 & 195 & $-101$ & 605\\
\end{tabular}
\end{ruledtabular}
\end{table}

The parameters $T_H$ and $T_Q$ define the boundaries of a mixed phase region
separating the hadronic and quark--gluon phases. The critical
temperature~$T_c$ as defined by lattice calculations should lie
between $T_H$ and $T_Q$\,, i.e. $T_c\simeq (T_H+T_Q)/2$\,. Earlier
lattice calculations (see e.g.~Ref.~\cite{Kar01}) predicted the values
\mbox{$T_c=(170\pm 10)$\,MeV} for
the baryon--free two--flavor QCD matter.
However, a noticeably larger value \mbox{$T_c=(192\pm 11)$\,MeV} was reported
recently in Ref.~\cite{Che06}. To probe sensitivity to the actual position of
the phase transition, we consider two EOS\hsp s with different $T_H$ and $T_Q$\,
(see Table~\ref{tab1}). The EOS--I corresponds to
$T_H=165$\,MeV and the parameters $\epsilon_H,\,\epsilon_Q$ used
in the parametrization LH12 of Ref.~\cite{Tea01}. In the EOS--II we choose
$T_H=190$\,MeV and scale $\epsilon_H,\,\epsilon_Q$ to get the same
values of $\epsilon/T^4$ as a function of~$T/T_H$\hsp\footnote
{
This is achieved by choosing the same $\epsilon_i/T_H^4\ (i=H,Q)$ for
these two EOSs.
}.
Finally, the parameters $B_Q, T_Q$ are found from the continuity
conditions for $P$ and $T$\,.

\begin{figure*}[htb!]
\centerline{\includegraphics[width=0.8\textwidth]{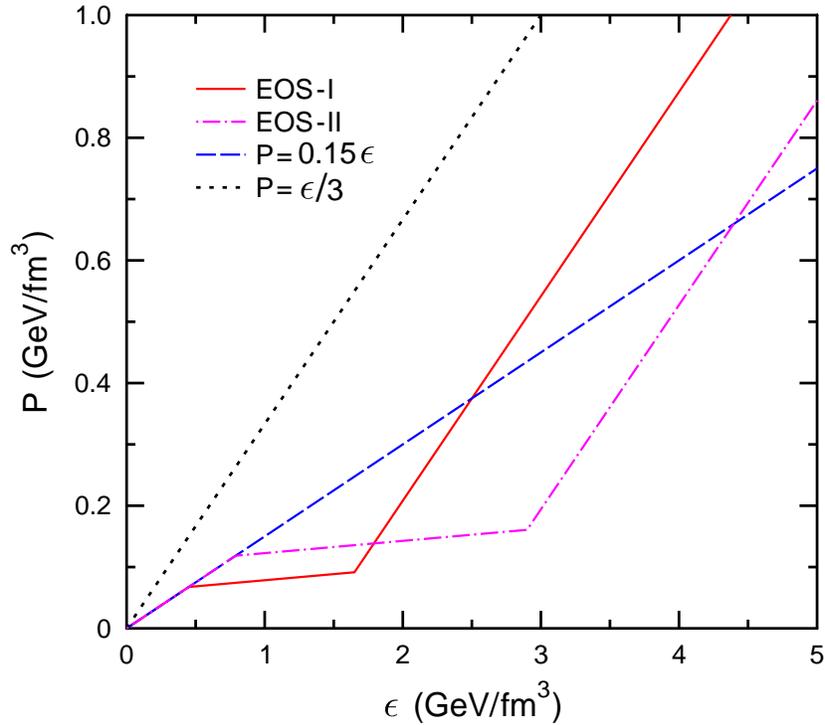}}
\caption{
Comparison of different EOS\hsp s used in this paper. The solid (dashed--dotted)
line is calculated using Eqs.~(\ref{heos})--(\ref{qeos}) with parameters
given by the set EOS--I (II) in Table~\ref{tab1}. The dashed and dotted
lines correspond to the hadronic EOS\hsp s with constant
$c^2_s=0.15$ and 1/3, respectively.
}
\label{fig1}
\end{figure*}

Unless stated otherwise, these EOS\hsp s are used in the calculations
presented in this paper. For comparison,
we have performed also calculations with several purely hadronic
EOS\hsp s. In this case we extend~\re{heos} to energy densities
$\epsilon>\epsilon_H$ with the same $\epsilon_H, T_H$ as in Table~\ref{tab1},
but choosing different $c_H^2=0.15$ and 1/3.
In Fig.~\ref{fig1} we compare the EOS--I and EOS--II
as well as two purely hadronic EOS\hsp s with constant sound
velocities $c_s=c_H$\,.
One can see that the mixed phase region in the EOS--II occupies larger interval
of energy densities, i.e. this EOS has a larger latent heat, $\epsilon_Q-\epsilon_H$,
as compared to the EOS--I. By this reason, the
life--times of the mixed phase will be longer for the EOS--II,
assuming the same initial state.

\subsection{Total energy and entropy}

Using the equations of fluid dynamics one can show that
the total energy and entropy of the fluid can be expressed as~\cite{Esk98}
\begin{eqnarray}
\label{tener}
&&E=\int\hspace*{-1mm} d\sigma_\mu T^{\mu 0}=
S_\perp\tau_0\int\limits_{-\infty}^{+\infty}\hspace*{-1mm}d\eta
\left[\epsilon\hsp\cosh{Y}\cosh{(\mbox{$Y-\eta$})}+
P\sinh{Y}\sinh{(Y-\eta)}\right]\hspace*{-1pt},\\
\label{tentr}
&&S=\int\hspace*{-1mm} d\sigma_\mu s U^{\mu}=
S_\perp\tau_0\int\limits_{-\infty}^{+\infty}
\hspace*{-1mm}d\eta\,s\hsp\cosh{(Y-\eta)}\,,
\end{eqnarray}
where $S_\perp$ is the transverse area of the fluid. The right-hand sides of
Eqs.~(\ref{tener})--(\ref{tentr}) give the values of the energy and entropy
at $\tau=\tau_0$\hsp .
Equations (\ref{tener}) and (\ref{tentr}) can be considered as sum
rules for the total energy and entropy of the produced particles.

Below we use~\re{tener} to constrain possible values of the parameters
characterizing the initial state. This is possible since the
total energy of produced particles can be estimated from experimental
data. Indeed, the value of the total energy loss, $\Delta
E=73\pm 6$ GeV per nucleon, has been obtained from the
the net baryon rapidity distribution in most central
Au+Au collisions~\cite{Bea04}. This gives the estimate of the total
energy of secondaries in the considered reaction:
\bel{teval}
E=N_{\rm part}\hsp\hsp\Delta E\simeq 26.1\,\textrm{TeV}\,,
\ee
where $N_{\rm part}\simeq 357$ is the mean number of
participating nucleons. Substituting the parametrization (\ref{incd})
into~\re{tener} and taking the value of $E$ from~\re{teval}, one gets
the relation between the parameters~$\epsilon_0, \eta_0, \sigma$\,.

We have checked that our numerical
code conserves the total energy $E$ and entropy $S$ at any hypersurface
$\sigma_\mu$ lying above the initial hyperbola $\tau=\tau_0$\,,
on the level better than~1\% up to very long times,
\mbox{$\tau\sim 10^3$\,fm/{\it c}}.

\subsection{Particle spectra at freeze--out}

The momentum spectra of secondary hadrons are calculated by applying the
standard Cooper--Frye formula~\cite{Coo74}, assuming that particles are emitted
without further rescatterings from the elements $d\sigma_\mu$ of the
freeze--out hypersurface $\tau=\tau_F(\eta)$\hsp . Then,
the invariant momentum distribution for each particle
species is given by the expression
\bel{spec}
{E}\frac{d^3N}{\hspace*{-4pt}d^3p}=\frac{d^3N}{dy\hsp d^2p_T}=
\frac{g}{(2\pi)^3}
\int d\sigma_\mu\hsp p^{\hsp\mu}\left\{\exp\left(\frac{p_\nu U^\nu_F-\mu_F}
{T_F}\right)\pm 1\right\}^{-1},
\ee
where $p^{\hsp\mu}$ is the 4--momentum of the particle, $y$ and $\bm{p}_T$
are, respectively, its longitudinal rapidity and transverse momentum,
$g$ denotes the particle's statistical weight. The subscript $F$ in the
collective 4--velocity $U^\mu$\,, temperature $T$ and chemical potential
$\mu$ implies that these quantities are taken on the freeze--out
hypersurface\footnote{
Below we assume that the chemical and
thermal freeze--out hypersurfaces coincide. In this case $\mu_F=0$ for
baryon--free matter.
}.
The plus or minus sign in the right-hand side of~\re{spec} correspond to fermions or
bosons, respectively.

As has been already stated, the effects of transverse expansion are
disregarded in our approach. Due to this reason, we cannot describe realistically
the $p_T$ spectra of produced hadrons, and analyze below
only the rapidity spectra.
For a cylindrical fireball with transverse cross section $S_\perp$
expanding only in the
longitudinal direction, one can write
$d\sigma^\mu=S_\perp\hsp (dz, \bm{0}, dt)^\mu$\,.
Using~\re{lcva} one arrives at the following relation
\bel{free}
d\sigma_\mu p^{\hsp\mu}=S_\perp\hsp m_T \left\{\tau_F (\eta)\cosh\hsp (y-\eta)-
\tau^\prime_F (\eta)\sinh\hsp (y-\eta)\right\} d\eta\,.
\ee
Here $m_T$ is the particle's transverse mass defined as
$m_T=\sqrt{m^2+\bm{p}_T^2}$\,, where~$m$ is the corresponding vacuum
mass.  In the same approximation one can also write the expression
\bel{expp}
p_\nu U^\nu_F=m_T \cosh\hsp (y-Y_F(\eta))\,,
\ee
where $Y_F(\eta)=Y\hsp (\tau_F\hsp (\eta), \eta)$\,.
An explicit expression for particle spectra at freeze--out
is obtained after substituting~(\ref{free})--(\ref{expp})
into~\re{spec} and integrating
over~$\eta$ from $-\infty$ to $+\infty$\,. Note that
Bjorken's model~\cite{Bjo83} corresponds to $Y_F=\eta$ and $\tau_F,
T_F$ independent of $\eta$. As can be seen from
Eqs.~(\ref{spec})--(\ref{expp}),
the rapidity distributions of all particles should be flat in this case.

We adopt a freeze--out criterion, assuming that a given fluid element
decouples from the rest of the fluid when its temperature decreases
below a~certain value~$T_F$\,. For finite--size initial conditions,
\mbox{$T(\tau_0,\eta)\to 0$} at \mbox{$|\eta|\to\infty$},
so that the fluid elements
at large $|\eta|$ have temperatures below $T_F$ from the very
beginning, i.e. at $\tau=\tau_0$. We treat these elements as decoupled
instantaneously~($\tau_F=\tau_0$) and use in~\re{spec} the initial
values of $Y$ and~$T$ instead of $Y_F$ and $T_F$\,. Direct
calculation shows, that such elements contribute only little to the
tails of the rapidity distributions.  The value of $T_F$ is considered
as an adjustable model parameter which is found from the best fit to
experimental data.

\subsection{Feeding from resonance decays}

In calculating particle spectra one should
take into account not only directly produced
particles but also feeding from resonance decays.
Below we assume that the freeze--out temperatures
for directly produced particles and corresponding resonances
are the same. One of the most important contributions
to the pion yield is given by $\rho\hsp (770)$--mesons.
The spectrum of $\pi^+$--mesons originating from these decays
is calculated by using the expression~\cite{Sol90}
\bel{rdc}
E_\pi\frac{d^3N_{\rho\to\pi^+}}{d^3 p\hspace*{3pt}}=
\frac{1}{3\pi}\int\limits_{2m_\pi}^\infty
\frac{d\hsp m_R\hsp w\hsp (m_R)}
{\sqrt{m_R^2-4m_\pi^2}}\int d^3p_R\hsp\frac{d^3N_R}{d^3p_R}
\,\delta\left(\frac{p\hsp p_R}{m_R}-\frac{m_R}{2}\right),
\ee
where the first integration corresponds to averaging over the mass
spectrum of \mbox{$\rho$--mesons}, $p_R$~and $p$ are, respectively, the
4--momenta of the $\rho$--resonance and of the secondary pion. The
normalization coefficient in~\re{rdc} takes into account that the
number of $\pi^+$--mesons produced in $\rho$--decays equals 2/3 of the
total multiplicity of $\rho$--mesons.  The freeze--out momentum
spectrum of $\rho$--mesons, $d^3N_R/d^3p_R$\,, is calculated using
Eqs.~(\ref{spec})--(\ref{expp}) with \mbox{$m=m_R$},
\mbox{$g=g_\rho=9$}. We use the~pa\-ra\-metrization of the
$\rho$--meson mass distribution, $w\hsp (m_R)$\,, sugges\-ted in
Ref.~\cite{Sol90}.

The feeding of the pion yields from other meson and baryon resonances
(\mbox{$R=\eta,$} $\omega, K^*, \Delta\ldots$) is obtained in the zero--width
approximation, assuming that the contribution of the resonance $R$ is
proportional to its equilibrium density $n_R\hsp(T_F)$, multiplied by
a factor $d_R$, the average number of $\pi^+$ mesons produced
in this resonance decay
($d_\rho=2/3, \mbox{$d_\eta=0.65\ldots $}$). The details of $n_R$ and
$d_R$ calculations can be found in Ref.~\cite{Beb92}. We have checked
for several resonances with two--body decays (e.g. for $R=K^*$)
that such a procedure yields a very good accuracy. As a result, we
get the following formula for the total resonance contribution to the
spectrum of $\pi^+$ mesons:
\bel{resc}
\sum\limits_R\frac{\ds d^3N_{R\to\pi^+}}{\ds dy\hsp d^2p_T}=
\alpha\hsp\frac{\ds d^3N_{\rho\to\pi^+}}
{\ds dy\hsp d^2p_T}\,,
\ee
where the enhancement factor $\alpha$ is defined as follows
\bel{enf}
\alpha=\sum\limits_R\frac{d_R}{d_\rho}\frac{n_R(T_F)}{n_\rho(T_F)}\,.
\ee
We include meson (baryon and antibaryon) resonances
with masses up to 1.3 (1.65) GeV and widths $\Gamma<150$ MeV.
The statistical weights, masses and branching ratios of these
resonances are taken from Ref.~\cite{PDG04}\,.
The factor $\alpha$ decreases gradually with decreasing freeze--out temperature:
\mbox{$\alpha=2.8, 2.4, 2.3$} for $T_F=165, 130, 100$\,MeV, respectively.

When calculating the kaon spectra we explicitly include feeding from
decays of $K^*(892)$ (in the zero--width approximation).
Higher resonances ($R=\phi, K_1\ldots$) are taken into account
by applying the same
procedure as for pions.  In this case the enhancement factor changes
from 1.5 to 1.2 when $T_F$ goes from 165 to 100  MeV.

\section{Results}

\subsection{Best fits of rapidity spectra}

Below we show the results for rapidity distributions of $\pi$-- and
$K$--mesons as well as antiprotons produced in central Au+Au
collisions at $\sqrt{s_{NN}}=200$\,GeV. In all calculations we use the
fireball radius $R=6.5$\,fm and $S_\perp=\pi R^2\simeq 133$\,fm$^2$.
The results are compared with data
of the BRAHMS Collaboration~\cite{Bea04,Bea05} for 5\% most central
collisions.

We have considered different profiles of the initial energy density,
ranging from the Gaussian--like ($\eta_0=0$) to the table--like
($\sigma=0$). We found that in the case of EOS--I it is not possible
to reproduce the BRAHMS data on the pion and kaon rapidity spectra
in Au+Au collisions
by choosing either too small ($\epsilon_0\lesssim 5$\,GeV/fm$^3$) or
too large ($\epsilon_0\gtrsim 15$\,GeV/fm$^3$) initial energy
densities. For such $\epsilon_0$ values the pion and kaon yields can not be
reproduced with any $T_F$\,. It is also found that the quality of fits
is noticeably reduced for initial energy density profiles with sharp
edges, corresponding to $\sigma<1$\,. As follows from the
constraint~(\ref{teval}), such profiles should have either
very large $\epsilon_0$ or a wide
plateau $-\eta_0<\eta<\eta_0$. This would lead to more flat rapidity
distributions of pions and kaons as compared to the BRAHMS data.

A few parameter sets which give good fits
with the EOS--I are listed in Table~\ref{tab2}.
All three sets from Table~\ref{tab2} give very similar rapidity
distributions for both pions and kaons. In these calculations we choose various
$\epsilon_0$ and $\sigma$ and determine $\eta_0$ from the total
energy constraint~(\ref{teval}).
\begin{table}[htb!]
\caption{\label{tab2}
Parameters of the initial states which give the
best fits of the pion, kaon and antiproton rapidity spectra observed
in central Au+Au collisions at $\sqrt{s_{NN}}=200$\,GeV.
All sets correspond to the EOS--I.
$T_0$~denotes the maximum temperature at $\tau=\tau_0$\,.
$E_1$ and $E_3$ are total energies of produced particles
within the rapidity intervals $|y|<1$ and $|y|<3$, respectively.}
\vspace*{2mm}
\begin{ruledtabular}
\begin{tabular}{ccllcccc}
set& $\epsilon_0$\,(GeV/fm$^3$) &
$\sigma$ & $\eta_0$ & $T_0$\,(MeV) &
$E_1$\,(TeV) & $E_3$\,(TeV)& $E/S$\,(GeV)\\
\hline
A & 10 & 1.74 & 0    & 279 & 1.53 & 9.25 & 0.89\\
B & 9  & 1.50 & 0.62 & 271 & 1.54 & 9.59 & 0.86\\
C & 8  & 1.30 & 1.14 & 263 & 1.49 & 9.55 & 0.86\\
\end{tabular}
\end{ruledtabular}
\end{table}
It is interesting that the initial
states A--C have approximately the same total entropy
$S\simeq 3\times 10^4$\,. This, in fact, should follow from the correct
description of total pion and kaon multiplicities\footnote
{
According to the Landau model~\cite{Lan53}, the
multiplicity of produced particles is proportional to the
total entropy.
}.
As one can see from the last column of Table~\ref{tab2},
the corresponding $E/S$--ratios fall into a narrow
interval $0.86-0.89$\,GeV. This observation is similar to the result of
Ref.~\cite{Cle98} that the observed ratio of the rest frame energy
to the multiplicity of produced hadrons
is constant as a function of the bombarding energy.

To check the sensitivity to the parameters of the phase transition,
we also calculate the pion and kaon rapidity distributions for the EOS--II.
It is found that with the same initial energy profiles as for the EOS--I
it is not possible to reproduce the observed spectra at any freeze--out
temperature. In particular, the predicted kaon yield is strongly
overestimated\hsp\footnote
{
We have checked that at fixed $T_F$ and the same initial conditions the freeze--out
times predicted for the EOS--II are noticeably longer
than for the EOS--I.
}
at \mbox{$100\,\textrm{MeV}<T_F<T_H=190$\,MeV}.
Nevertheless, the BRAHMS data can be well reproduced with the EOS--II too
when taking smaller initial energy densities as compared with the EOS--I.
Fits of approximately same quality are obtained for $\epsilon_0\simeq 5$\,GeV/fm$^3$\,.
As before, in choosing the initial conditions we apply the constraint~(\ref{teval})
for the total energy of produced particles.
Similarly to the case of the EOS--I, the data are better reproduced for initial
profiles with small $\eta_0\lesssim 1$\,.

\begin{figure*}[htb!]
\centerline{\includegraphics[width=0.8\textwidth]{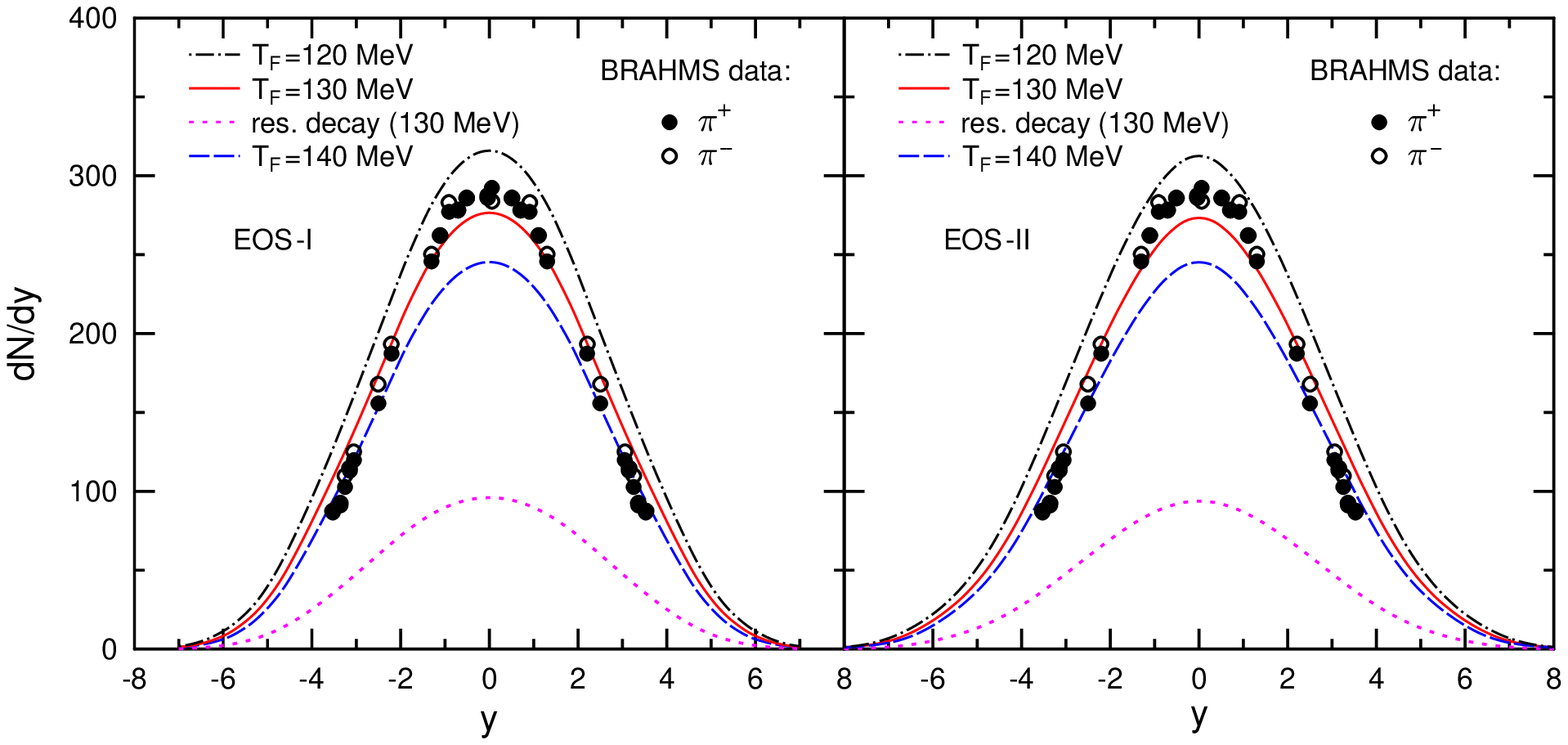}}
\caption{Rapidity distribution of $\pi^+$--mesons in central Au+Au
collisions at $\sqrt{s_{NN}}=200$\,GeV.
Left panel shows results of hydrodynamical calculations for the EOS--I
and initial conditions (\ref{incd}) with the parameters
\mbox{$\epsilon_0=10$\,GeV/fm$^3$, $\eta_0=0$, $\sigma=1.74$} (set A from
Table~\ref{tab2}).
Right panel corresponds to the EOS--II and the parameters
\mbox{$\epsilon_0=5$\,GeV/fm$^3$, $\eta_0=0$, $\sigma=2.02$}.
Solid, dashed and dashed--dotted curves are calculated for different values of
the freeze--out temperature $T_F$\,. The dotted lines show contributions of
resonance decays in the case $T_F=130$\,MeV. Experimental data are
taken from Ref.~\cite{Bea05}.
}
\label{fig2}
\end{figure*}

\begin{figure*}[htb!]
\centerline{\includegraphics[width=0.8\textwidth]{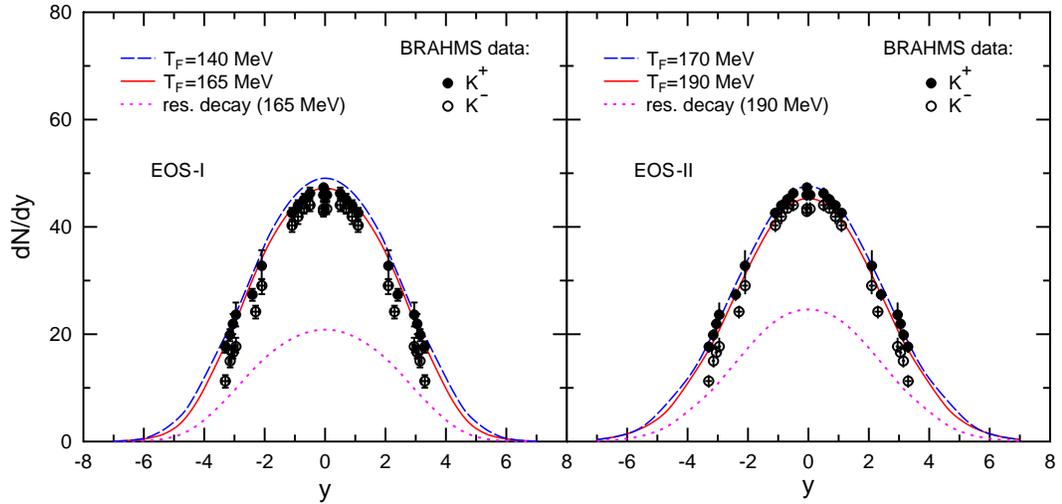}}
\caption{Same as Fig.~\ref{fig2}, but for $K^+$ rapidity distributions.}
\label{fig3}
\end{figure*}

Figures~\ref{fig2}--\ref{fig3} show the model results
for pion and kaon rapidity distributions
obtained for the EOS--I and EOS--II. These results correspond to
Gaussian initial profiles
with $\eta_0=0$\,. For both EOS\hsp s we choose the parameter $\epsilon_0$
to obtain the best fit of the BRAHMS data\footnote
{
We did not try to achieve a perfect fit
of these data, bearing in mind that their systematic
errors are quite big, about 15\% in the rapidity region
$|y|>1.3$~\cite{Bea05}.
}.
Although the overall fits are very similar for both EOS\hsp s,
the rapidity spectra obtained with the EOS--II are slightly broader
than those with the EOS--I.
In the same figures we demonstrate sensitivity to the choice of
the freeze--out temperature.
The best fits of the pion spectrum for EOS--I and EOS--II are
achieved with $T_F\simeq 130$\,MeV (see Fig.~\ref{fig2}).
On the other hand, the kaon spectrum can be well reproduced
only by assuming that kaons decouple at the very beginning of the hadronic stage,
i.e. at \mbox{$T_F\simeq T_H=165\,(190)$\,MeV} for EOS--I (II). The contribution
of resonance decays turns out to be rather significant, especially in the central
rapidity region, where it amounts to about 35\% (45\%) of the total
pion (kaon) yield.

According to Fig.~\ref{fig2}, larger yields
of secondary pions are predicted for smaller freeze--out temperatures. A much weaker
sensitivity to $T_F$ is found for kaons (see Fig.~\ref{fig3}). This difference can be
explained by the large difference between the pion and kaon masses.
Indeed, in the case of direct pions,
a good approximation at $T_F>100$\,MeV  is to
replace the transverse mass $m_T$ in Eqs.~(\ref{spec})--(\ref{expp}) by
the pion transverse momentum~$p_T$. Neglecting the second term
in the right-hand side of~\re{free}, one
can show that the rapidity distribution of pions at $y=0$ is
proportional to
$\xi=\tau_F(\eta)\hsp\cosh{\eta}\cdot T_F^{\hsp 3}/\cosh^3{Y_F(\eta)}$
integrated over all $\eta$\,. For a
rough estimate, one can use the Bjorken relations~\cite{Bjo83}
$\mbox{$Y_F=\eta$},\, s_F\tau_F=s_0\tau_0$\,, where $s_F$ is the entropy density
at $T=T_F$\,. Using~\re{heos} one gets
\mbox{$\tau_F\propto s_F^{-1}\propto T_F^{-1/c_H^2}$} and therefore,
$\xi\propto T_F^{3-1/c_H^2}$\,.
This shows that for $c_H^2<1/3$ the pion yield grows with decreasing
$T_F$. Qualitatively, one can say that at low enough
$c_H$  the increase of the spatial volume at freeze--out compensates
for the decrease of the pion occupation numbers at smaller $T_F$. This
effect is somewhat reduced because of a decreasing resonance contribution
at smaller temperatures. It is obvious that for kaons this effect
should be much weaker due to the presence of the activation exponent
$\exp{(-m_K/T_F)}$\,. In fact, a numerical calculation for the same EOS
and initial state shows that the kaon yield changes
nonmonotonically: first it slightly increases when temperature goes
down but then it starts to decrease at
lower~$T_F$\,.

To study sensitivity of particle spectra to the presence of
the phase transition, we have performed calculations
with purely hadronic EOS\hsp s. In this case we
use the same initial conditions as before and apply
\re{heos} for all stages of the reaction, including
high density states. Our analysis shows that for soft hadronic EOS
with $c_H^2\simeq 0.1-0.2$ it is possible to reproduce the observed pion
and kaon data with approximately the same fit quality as in
the calculations with the quark--gluon phase transition. Furthermore, the
corresponding freeze--out temperatures do not change significantly.
\begin{figure*}[htb!]
\vspace*{-6mm}
\centerline{\includegraphics[width=0.8\textwidth]{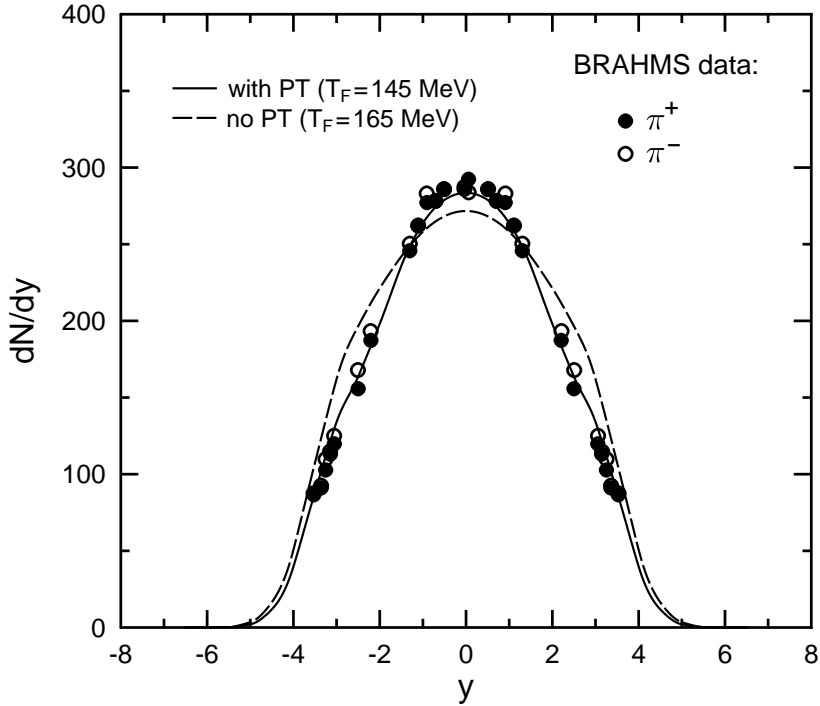}}
\vspace*{-4mm}
\caption{
Rapidity distributions of $\pi^+$--mesons in central Au+Au
collisions at \mbox{$\sqrt{s_{NN}}=200$} GeV calculated for the initial condition with
the parameters \mbox{$\epsilon_0=50$\,GeV/fm$^3$,} $\eta_0=0$, $\sigma=0.96$.
Solid line corresponds to the EOS with the phase transition (see the text).
The dashed line is calculated the purely hadronic EOS.
In both cases \mbox{$c_H^2=1/3$}.
Experimental data are taken from Ref.~\cite{Bea05}.
}
\label{fig4}
\end{figure*}
However, we could not achieve satisfactory fits for the ''hard'' hadronic EOS with
$c_H^2\geqslant 1/3$. This is demonstrated in Fig.~\ref{fig4}
where we compare calculations for two EOS\hsp s
with and without phase transition. In both calculations~$c_H^2=1/3$.
In the first case with we use Eqs.~\mbox{(\ref{heos})--(\ref{qeos})}
with the same $T_H,\,\epsilon_H$ and $\epsilon_Q$ as
for the EOS--I, but choose $c_H^2=1/3$\,.
The hadronic EOS is obtained by extending \re{heos} to all energy densities.
We have found that calculations with $c_H^2=1/3$ require much higher
initial energy densities as compared to the EOS with $c_H^2=0.15$\,.
One can see that this hadronic EOS predicts a too wide pion rapidity
distribution. The same conclusion is valid for kaons. The reason is that the higher
pressure gives a stronger push to the matter in forward and backward
directions. From these findings we conclude that a certain degree of softening
of the EOS is required to  reproduce the pion and kaon
rapidity distributions.

\begin{figure*}[htb!]
\vspace*{-6mm}
\centerline{\includegraphics[width=0.8\textwidth]{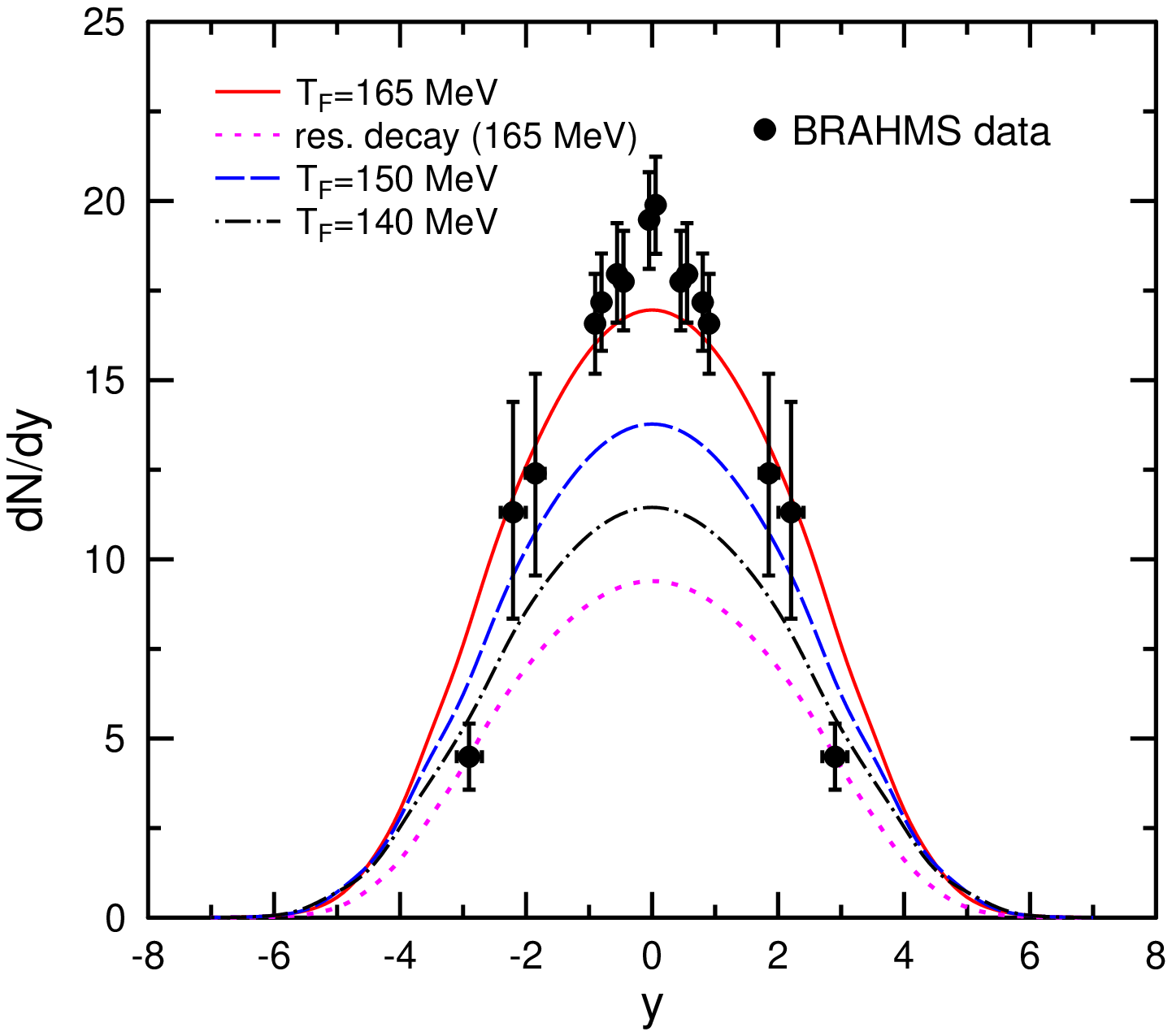}}
\vspace*{-4mm}
\caption{
Rapidity distributions of antiprotons in central Au+Au
collisions at \mbox{$\sqrt{s_{NN}}=200$\,GeV}.
Shown are results of hydrodynamical calculations
for the EOS--I and the parameter set A.
All results are obtained assuming $\mu_F=0$.
Experimental data are taken from Ref.~\cite{Bea04}.
}
\label{fig5}
\end{figure*}

It turns out that our model
can also reproduce reasonably well the antiproton rapidity spectra
measured by the BRAHMS Collaboration~\cite{Bea04}.
Figure~\ref{fig5} shows the antiproton rapidity distributions,
calculated for the EOS--I and the parameter set A. In this case we explicitly take
into account the contribution of the
\mbox{$\ov{\Delta}(1232)\to\pi\ov{p}$} decays, ignoring the
width of \mbox{$\Delta$--isobars}. Contributions of higher
antibaryon resonances are taken into account in a similar way as for
pions and kaons. The resonance contribution is about 55\% at
$T_F=165$\,MeV. One should consider these results as an upper
bound for the antiproton yield. A more realistic model should include
effects of the nonzero baryon chemical potential which will certainly reduce
the antibaryon yield. The thermal model
analysis of RHIC data, performed in Refs.~\cite{And04,And06}, gives rather low
values for the baryon chemical potentials, $\mu_F\sim 30$\,MeV,
at midrapidity. This will suppress the antiproton yield by
a factor $\sim\exp{(-\mu_F/T_F)}\sim 0.8$\,.

\subsection{Rapidity distribution of total energy}

We have calculated additionally
the rapidity distribution of the total energy of secondary
particles, $dE/dy$\,,
in order to check the energy balance in the considered reaction.
In this calculation we take into account not only
direct pions and kaons (charged and neutral), but also heavier
mesons and $B\ov{B}$ pairs (the same set of resonances as in
the calculation of pion and kaon spectra). The contribution of heavy
mesons and $B\ov{B}$ pairs was calculated in the zero--width
approximation at the temperature $T_F=165$\,MeV.  By integrating
$dE/dy$, we have determined $E_1$ and $E_3$, the total energies of secondaries within
the rapidity intervals $|y|<1$ and $|y|<3$, respectively.
The BRAHMS Collaboration estimated $E_{1,3}$
from the rapidity distributions of charged pions, kaons, protons and
antiprotons in most central Au+Au collisions at
$\sqrt{s_{NN}}=200$\,GeV. The values
$E_1\simeq 1.5\,\textrm{TeV},\hspace*{2ex}\mbox{$E_3\simeq 9\,\textrm{TeV}$}$
have been reported in Ref.~\cite{Ars05}.
From Table~\ref{tab2} one can see that these values are well reproduced by
the model.

Based on the above analysis we conclude that
within the hydrodynamical model the BRAHMS data
can be well described with the EOS-I and EOS--II
and the parameters of the initial state ($\tau_0=1$\,fm/{\it c})
$\sigma\simeq 1.5-2,~~\eta_0\lesssim 1$\,.
The maximal initial energy density, $\epsilon_0$, is sensitive
to the critical temperature of the phase transition.
For the EOS--I ($T_c\simeq 167$\,MeV) we get the
estimate $\epsilon_0\simeq 9\pm 1$\,\textrm{GeV/fm}$^3$
while for the EOS--II ($T_c\simeq 192$\,MeV) the required
values of $\epsilon_0$ are lower by about a factor of two.

These profiles are intermediate between the Landau and Bjorken
limits. It is worth noting that the observed pion rapidity distribution
can be well approximated by the Gaussian with the width
$\sigma_{\rm exp}\simeq 2.3$~\cite{Bea05}.
According to the Landau model, the width of the distribution
is given by the expression~\cite{Shu72}
\bel{lwid}
\sigma_{\rm Lan}^2\simeq \frac{8}{3}\frac{\ds c_s^2}
{\ds 1-c_s^4}\ln{\frac{\sqrt{s_{NN}}}{2\hsp m_N}}\,,
\ee
where $m_N$ is the nucleon mass. For $c_s^2=1/3$ this gives
$\sigma_{\rm Lan}\simeq 2.16$\,, the value often quoted in the
literature (see e.g.~\cite{Bea05}). On the other hand,
for $c_s^2=0.15$ (which is preferable within our model) the width
is only 1.38 i.e. noticeably smaller than observed by the
BRAHMS Collaboration. This shows that deviations from the simple Landau
model are rather significant.

\subsection{Dynamical evolution of matter}

\begin{figure*}[htb!]
\hspace*{-12mm}\includegraphics[width=0.9\textwidth]{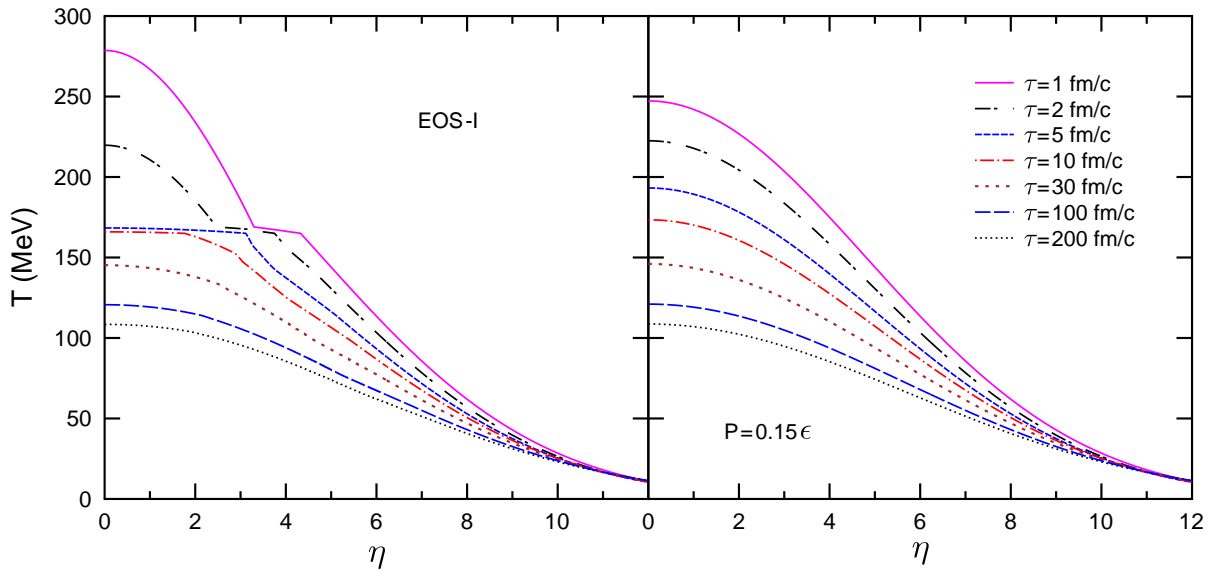}
\caption{
Time dependence of temperature as a function of $\eta$
calculated for the parameter set A (only the forward
hemisphere is shown). Left and right panels correspond, respectively,
to the EOS--I and the hadronic EOS $P=c_H^2\hsp\epsilon$\, with
$c_H^2=0.15$\,.
}
\label{fig6}
\end{figure*}
\begin{figure*}[htb!]
\hspace*{-12mm}\includegraphics[width=0.9\textwidth]{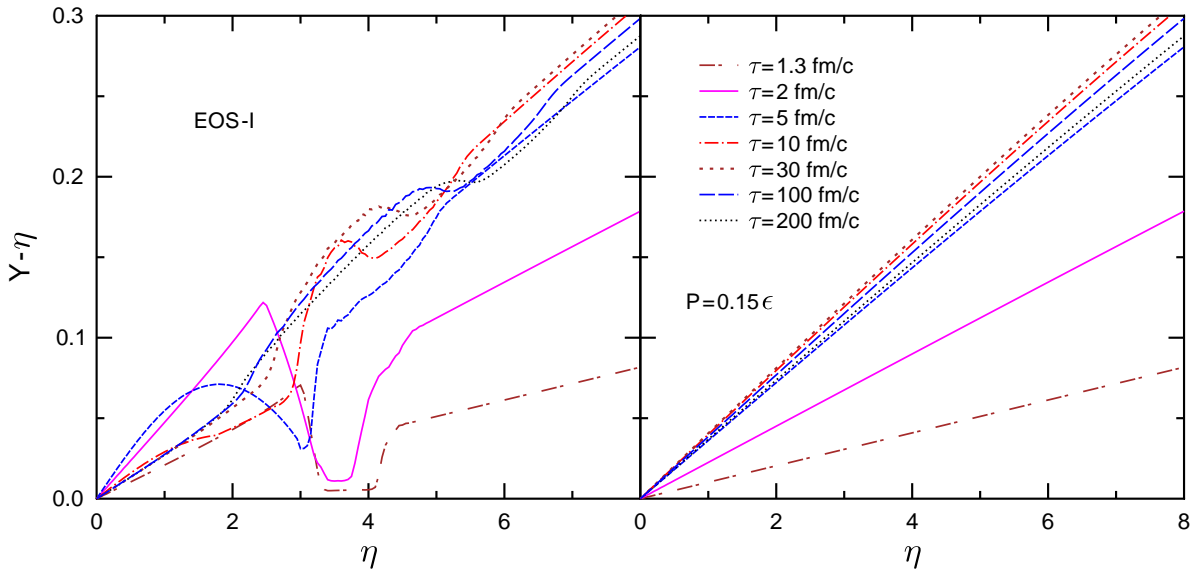}
\caption{
Same as Fig.~\ref{fig6}, but for collective rapidity profiles.
}
\label{fig7}
\end{figure*}

Finally, after we have determined the initial conditions
which lead to a reasonable description of the observed rapidity spectra,
we can use the strength of the hydrodynamical model
to follow the dynamical evolution of matter.
Below we present the results for two EOS\hsp s, with and without
the phase transition, for the parameter set A.
Figures~\ref{fig6}--\ref{fig7} show profiles of the temperature
and the collective rapidity at different proper times $\tau$.
The main difference is that in the case of phase transition the model
predicts appearance
of a flat shoulder in $T(\eta)$ and local minima in $Y(\eta)$\,, which are clearly seen
at $\tau\lesssim 10$\,fm/{\it c}\,. This is a consequence of the mixed phase
which has a life time $\Delta\tau\sim 10$\,fm/{\it c}\,. According to
Figs.~\ref{fig6}--\ref{fig7}, the ''memory'' of the quark--gluon phase
is practically washed out at $\tau\gtrsim 30$\,fm/{\it c}\,.
As one can see from Fig.~\ref{fig7}, at such times deviations
from the Bjorken scaling~($Y=\eta$) do not exceed 5\%.

\begin{figure*}[htb!]
\hspace*{-12mm}\includegraphics[width=0.9\textwidth]{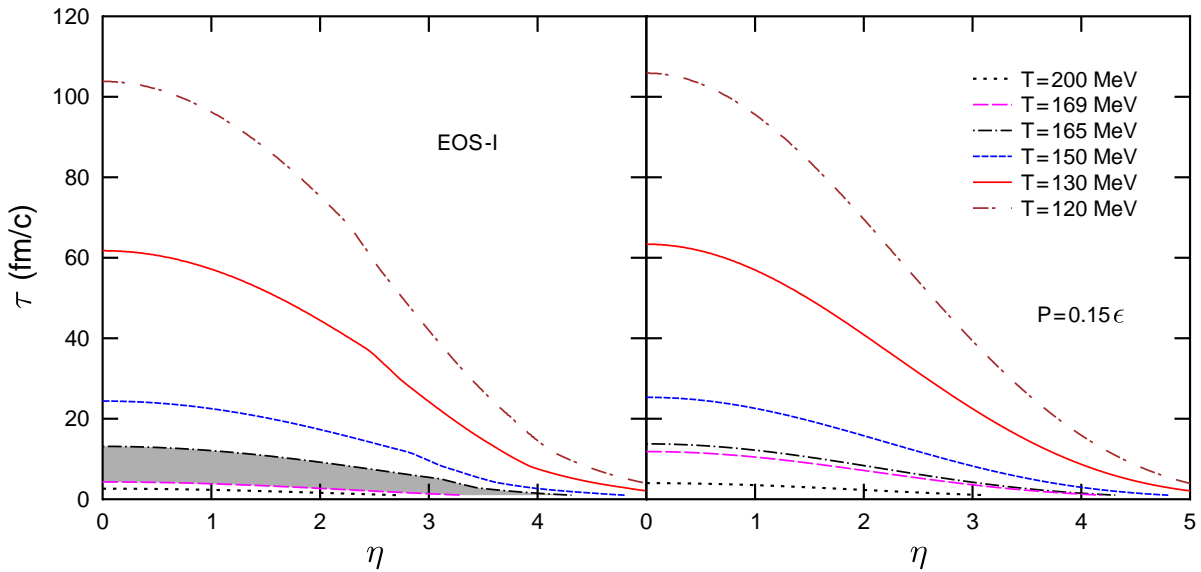}
\caption{
Isotherms in the $\eta-\tau$ plane
calculated for the parameter set A.
Left and right panel corresponds to the same
EOS\hsp s as in Fig.~\ref{fig6}\,.
Shaded region indicates the mixed phase.
}
\label{fig8}
\end{figure*}

Figure~\ref{fig8} shows the matter isotherms in the
$\eta-\tau$ plane. One can clearly see that the initial stage of the
evolution, when matter is in the quark--gluon phase, lasts only for a very short
time, of about 5 fm/{\it c}\,.
The region of the mixed phase is crossed in
less than 10 fm/{\it c}\,. This clearly shows that the slowing down of expansion
associated with the ''soft point'' of the EOS plays no role, when the
initial state lies much higher in energy density than the phase
transition region. In this situation the system spends the longest
time in the hadronic phase and the late evolution
is not sensitive to the phase transition. The freeze-out at $T_F=130$\,MeV requires
an expansion time of about 60 fm/{\it c}  at $\eta=0$. This is certainly a
very long time which is apparently in contradiction with experimental
findings. Indeed, the interferometric measurements~\cite{Adl01}
show much shorter times of hadron emission, of the order of
10 fm/{\it c}\,. This discrepancy can not be removed by considering other EOS
or initial conditions. A considerable reduction of the freeze--out times can be
achieved by including the effects of transverse expansion and chemical
nonequilibrium~\cite{Hir02}. However, this will not change
essentially the dynamics of the early stage ($\tau\lesssim 10$\,fm/{\it c})
when expansion is predominantly one--dimensional. A more radical solution
of the ''short time puzzle'' could be an explosive decomposition of the quark--gluon
plasma, proposed in Ref.~\cite{Mis99}. This may happen at very
early times, right after crossing the critical temperature line, when the
plasma pressure becomes rather small. We shall consider
this possibility in a forthcoming publication.

\section{Summary and discussion}

In this paper we have generalized Bjorken's scaling hydrodynamics for
finite--size profiles of energy density in pseudorapidity space.
The hydrodynamical equations were solved numerically
in $\tau-\eta$ coordinates
starting from the initial time $\tau_0=1$ fm/{\it c}  until the
freeze--out stage. The sensitivity of the final particle distributions to
the initial conditions, the freeze--out temperature and the EOS has been
investigated. A comparison of $\pi, K, \ov{p}$\ rapidity spectra with
the BRAHMS data for central Au+Au collisions at
$\sqrt{s_{NN}}=200$\,GeV has been made. Best agreement with these
data is obtained for initial states with nearly Gaussian profiles
of the energy density.
In choosing the initial conditions we impose the constraint on
the total energy of produced particles.
It is found that the maximum energy density of the initial state, $\epsilon_0$,
is sensitive to the parameters of a possible deconfinement transition.
The BRAHMS data are well reproduced with
$\epsilon_0$ of about 10 (5)~GeV/fm$^3$ for the critical temperature
$T_c\sim 165 \,(190)$\,MeV.
The only unsatisfactory aspect of these calculations is the prediction of
a very long freeze--out times, $\sim 50$\,fm/{\it c}  for pions.

We would like to comment on several points.

It is clear that our 1+1 dimensional model can not be valid at late stages
of heavy--ion collisions, and the transverse flow effects should be included
into a more realistic approach. On the other hand,
the above--mentioned 2+1 dimensional models
\mbox{\cite{Kol99,Bas00,Per00,Kol01,Tea01}}, which assume Bjorken
scaling in the beam direction, are apparently not accurate too, even
for the slice around \mbox{$\eta=0$}. Indeed, in contrast to the Bjorken
model, our calculations show that total entropy in different
pseudorapidity intervals does not stay constant during the expansion.
Due to the pressure gradients along the beam axis and
corresponding fluid's acceleration, the entropy is transferred from central
pseudorapidity bins to the periphery. For instance, the entropy
in the central bin $|\eta|<1$ drops by about 15\% during
the evolution. Therefore, only full 3D models can provide a more
reliable description.

It is interesting to note that the viscosity terms, omitted in this paper,
should lead to the opposite effects, namely to slower cooling
and smaller acceleration of the fluid~\cite{Tea04,Bai06}.
Therefore, to describe the observed data, we would need somewhat broader
initial energy density profiles and accordingly lower $\epsilon_0$ values.
In principle, our simple model can be used for a more quantitative study
of these effects.

We have performed calculations with the initial time $\tau_0=1$\,fm/{\it c}\,.
Of course, one can start the hydrodynamical evolution from an earlier
time, i.e. assuming smaller $\tau_0$\,. In this case one should choose
accordingly higher initial energy densities.  But $\tau_0$ cannot be
taken too small, since at very early times the energy is most likely
stored in strong chromofields~\cite{McL94}. The quark--gluon plasma is
produced as a result of the decay of these fields (see e.g. Ref.~\cite{Fuk06}
and references therein). Estimates show that
the characteristic decay times are in the range $0.3-1.0$~fm/{\it c}\,.  At
earlier times the system will contain both the fields as well as produced
partons, and the evolution equations will be more complicated, see e.g.
Refs.~\cite{Mis02,Mis06}.

It is obvious that the Cooper--Frye scenario of the freeze-out
process, applied in this paper is too
simplified. This was demonstrated e.g. in Ref.~\cite{Bra99}.
One should also have in mind that the freeze--out temperatures
obtained in our model will be modified by the
effects of transverse expansion and chemical nonequilibrium.
Attempts to achieve a more realistic
description of the freeze--out stage have been recently made in
Refs.~\cite{Bas00, Tea01} where a transport model was
applied to describe evolution of the hadronic phase.
In this approach the solution of fluid--dynamical equations is
used to obtain initial conditions for transport calculations
at later stages of a heavy--ion collision. We are planning to
use a similar approach in the future.

\begin{acknowledgments}
The authors thank I.G. Bearden, J.J. Gaardh{\o}je, M.I. Gorenstein,
Yu.B.~Ivanov, L.~McLerran, D.Yu.~Pe\-ressounko, D.H. Rischke, and
V.N. Russkikh for useful discussions.
This work was supported in part by the BMBF, GSI, the DFG grant~~436
RUS ~~~\mbox{113/711/0--2} (Germany) and the grants RFBR 05--02--04013 and
NS--8756.2006.2 (Russia).
\end{acknowledgments}



\vspace*{-3mm}
\begin{thebibliography}{00}

\bibitem{Lan53}
        L.D. Landau, Izv. Akad. Nauk Ser. Fiz. \textbf{17}, 51
        (1953); in Collected Papers of L.D. Landau,
        Pergamon, Oxford, 1965, p.~665.

\bibitem{Bjo83}
        J.D. Bjorken, Phys. Rev. D \textbf{27}, 140 (1983).

\bibitem{Mel58}
        G.A.~Milekhin,~Zh.~Eksp.~Teor.~Fiz.~\textbf{35},~1185 (1958)
        [Sov.~Phys.--JETP~\textbf{8},~829(1959)].

\bibitem{Tar77}
        Yu.A. Tarasov, Yad. Fiz. \textbf{26}, 770 (1977)
        [\mbox{Sov.~J.~Nucl.~Phys.} \textbf{26},~405 (1977)].

\bibitem{Mis83}
        I.N. Mishustin and L.M. Satarov, Yad.~Fiz. \textbf{37}, 894 (1983)
        [\mbox{Sov.~J.~Nucl.~Phys.}~\textbf{37}, 532 (1983)].

\bibitem{Bla87}
        J.P. Blaizot and J.Y. Ollitrault,
        Phys. Rev. D \textbf{36}, 916 (1987).

\bibitem{Esk98}
        K.J. Eskola, K. Kajantie, and P.V. Ruuskanen,
        Eur. Phys. J. C \textbf{1}, 627 (1998).

\bibitem{Moh03}
        B. Mohanty and J. Alam, Phys. Rev. C \textbf{68}, 064903 (2003).

\bibitem{Kol99}
         P.F. Kolb, J. Sollfrank, and U. Heinz, Phys. Lett.
         \textbf{B459}, 667 (1999).

\bibitem{Bas00}
        S.A. Bass and A. Dumitru, Phys. Rev. C \textbf{61}, 064909 (2000).

\bibitem{Per00}
         D.Yu. Peressounko and Yu.E. Pokrovsky,
         Nucl. Phys. \textbf{A669}, 196 (2000).

\bibitem{Kol01} P.F. Kolb, P. Huovinen, and U. Heinz,
         H. Heiselberg, Phys. Lett. \textbf{B500}, 232 (2001).

\bibitem{Tea01}
        D. Teaney, J. Lauret, and E.V. Shuryak,
        Phys. Rev. Lett. \textbf{86}, 4783 (2001);
        \mbox{nucl--th/0110037}.

\bibitem{Sto80} H. St\"ocker, J.A. Maruhn, and W.Greiner,
        Phys. Rev. Lett. \textbf{44}, 725 (1980).

\bibitem{Ris95a}
        D.H. Rischke, Y. P\"urs\"un, J.A. Maruhn, H. St\"ocker,
        and W. Greiner, Heavy Ion Phys. \textbf{1}, 309 (1995).

\bibitem{Non00}
        C. Nonaka, E. Honda, and S. Muroya,
        Eur. Phys. J. C \textbf{17}, 663 (2000).

\bibitem{Hir02}
        T. Hirano, Phys. Rev. C \textbf{65}, 011901(R) (2002);\\
        T. Hirano and K. Tsuda, Phys. Rev. C \textbf{66}, 054905 (2002).

\bibitem{Ham05}
        Y. Hama, T. Kodama, and O. Socolowski Jr.,
        Braz. J. Phys. \textbf{35}, 24 (2005).

\bibitem{Non06}
        C. Nonaka and S. Bass, Nucl. Phys. \textbf{A774}, 873 (2006).

\bibitem{Ams78}
        A.A. Amsden, A.S. Goldhaber, F.H. Harlow, and J.R. Nix,
        Phys. Rev.~C \textbf{17}, 2080 (1978).

\bibitem{Cla86}
        R.B. Clare and D. Strottman,
        Phys. Rep. \textbf{141}, 178 (1986).

\bibitem{Bar87}
        H.W. Barz, B. K\"ampfer, L.P. Csernai, and B. Lukacs,
        Nucl. Phys. \textbf{A465}, 743 (1987).

\bibitem{Mis88} I.N. Mishustin, V.N. Russkikh, and L.M.~Satarov,
        \mbox{Yad.~Fiz. \textbf{48}, 711 (1988)}
        [Sov.~J. Nucl. Phys. \textbf{48},~454 (1988)];
        Nucl. Phys. \textbf{A494}, 595 (1989).

\bibitem{Kat93}
        U. Katscher, D.H. Rischke, J.A. Maruhn, W. Greiner,
        I.N. Mishustin, and L.M.~Satarov, Z.~Phys. \textbf{A346}, 209 (1993).

\bibitem{Bra00}
        J. Brachmann, S. Soff, A. Dumitru, H. St\"ocker, J.A. Maruhn,
        W. Greiner, L.V.~Bravina, and D.H. Rischke,
        Phys. Rev. C \textbf{61}, 024909 (2000).

\bibitem{Ton03}
        Yu.B. Ivanov, V.N. Russkikh, and V.D. Toneev,
        Phys. Rev. C \textbf{73}, 044904 (2006).

\bibitem{Sto05}
         H.~St\"ocker, Nucl. Phys. \textbf{A750}, 121 (2005).

\bibitem{Bea04}
        BRAHMS Collaboration, I.G. Bearden {\it et~al.},
        Phys. Rev. Lett. \textbf{93}, 102301 (2004).

\bibitem{Bea05}
        BRAHMS Collaboration, I.G. Bearden {\it et~al.},
        Phys. Rev. Lett. \textbf{94}, 162301 (2005).

\bibitem{Bel96}
         M. Le Bellac, Thermal Field Theory, Cambridge University Press,
         Cambrige, 1996.

\bibitem{Ris95b}
        D.H. Rischke, S. Bernard, and J.A. Maruhn,
        Nucl. Phys. \textbf{A595}, 346 (1995).

\bibitem{Bor73}
        J.P. Boris and D.L. Book, J. Comp. Phys. \textbf{11}, 38 (1973).

\bibitem{Kar01}
        F. Karsch, E. Laermann, and A. Peikert,
        Nucl. Phys. \textbf{B605}, 579 (2001).

\bibitem{Che06}
        M. Cheng {\it et al.}, Phys. Rev. D \textbf{74}, 054507 (2006).

\bibitem{Coo74}
        F. Cooper and G. Frye, Phys. Rev. D \textbf{10}, 186 (1974).

\bibitem{Sol90}
        J. Sollfrank, P. Koch, and U.W. Heinz,
        Phys. Lett. \textbf{B252}, 256 (1990);\\
        Z. Phys. \textbf{C52}, 593 (1991).

\bibitem{Beb92}
        H. Bebie, P. Gerber, J.L. Goity, and H. Leutwyler,
        Nucl. Phys. \textbf{B378}, 95 (1992).

\bibitem{PDG04}
        Particle Data Group, S. Eidelman {\it et al.},
        Phys. Lett. \textbf{B592}, 1 (2004).

\bibitem{Cle98}
        J. Cleymans and K. Redlich, Phys. Rev. Lett. \textbf{81},
        5284 (1998).

\bibitem{And04}
        A. Andronic and P. Braun--Munzinger,
        Lect. Notes Phys. \textbf{652}, 35 (2004).

\bibitem{And06}
        A. Andronic, P. Braun--Munzinger, and J. Stachel,
        Nucl. Phys. \textbf{A772}, 167 (2006).

\bibitem{Ars05}
         BRAHMS Collaboration, J. Arsene {\it et al.},
         Nucl. Phys. \textbf{A757}, 1 (2005).

\bibitem{Shu72}
         E.V. Shuryak, Yad. Fiz. \textbf{16}, 395 (1972).

\bibitem{Adl01}
         STAR Collaboration, C. Adler {\it et al.},
         Phys. Rev. Lett. \textbf{87}, 82301 (2001).

\bibitem{Mis99}
        I.N. Mishustin, Phys. Rev. Lett. \textbf{82}, 4779 (1999).

\bibitem{Tea04}
        Derek A. Teaney, J. Phys. G: Nucl. Part. Phys. \textbf{30},
        S1247 (2004).

\bibitem{Bai06}
        Rudolf Baier and Paul Romatschke, nucl--th/0610108.

\bibitem{McL94}
        L. McLerran and R. Venugopalan, Phys. Rev. D \textbf{49}, 2233 (1994).

\bibitem{Fuk06}
        K. Fukushima, F. Gelis, and L. McLerran, hep--ph/0610416.

\bibitem{Mis02}
        I.N. Mishustin and J.I. Kapusta,
        Phys. Rev. Lett. \textbf{88}, 112501 (2002).

\bibitem{Mis06}
        Igor N. Mishustin and Konstantin A. Lyakhov, hep--ph/0612069.

\bibitem{Bra99}
        L.V. Bravina, I.N. Mishustin, J.P. Bondorf, A. Faessler, and
        E.E. Zabrodin, Phys. Rev. C \textbf{60}, 044095 (1999).

\end{thebibliography}
\end{document}